\documentclass{appolb}
\usepackage{epsfig}
\begin{document}
\pagestyle{plain}
\newcount\eLiNe\eLiNe=\inputlineno\advance\eLiNe by -1

\title{On the anomalous t-quark charge asymmetry and
 noncontractibility of the physical space}
\author{Davor Palle}
\address{Zavod za teorijsku fiziku, Institut Rugjer Bo\v skovi\' c \\
Bijeni\v cka cesta 54, 10000 Zagreb, Croatia}
\maketitle
\begin{abstract}
Heavy flavour production at hadron colliders represents a
very promising field to test perturbative QCD. The integrated
forward-backward asymmetry of the top-antitop quark production
is particularly sensitive to any deviation from the standard
QCD calculations. The two Tevatron collaborations, CDF and D0, 
reported a much larger t-quark charge asymmetry than predicted by the
theory. We show that the QCD in noncontractible space, 
where the minimal distance is fixed by weak interactions,
enhances the asymmetry by more than a factor of 3 (5) at
the parton level in leading order of the coupling
for the Tevatron (LHC) center of mass energies.
This result should not be a surprise since the asymmetry observable 
directly explores the far ultraviolet sector of the spacelike domain
of the Minkowski spacetime.
\end{abstract}

\section{Introduction and motivation}

The missing flavour-mixed light neutrinos, no lepton or baryon number
violation and the absence
of any candidate for a dark matter particle, call for a substantial
improvement of the Standard Model (SM)
of electroweak and strong interactions.
Namely, the dominance of the baryon over antibaryon matter in the Universe
suggests that baryon or lepton number be broken in particle physics
\cite{Sakharov}.
It is well known, for example, that the broken lepton 
number can induce the breaking of the baryon number \cite{Kolb}.
Any alternative to the presence of the cold dark matter in the Universe
supposes drastic changes not only of the General Relativity, but also
of the nonrelativistic Newtonian theory of gravity \cite{Peacock}.
However, the direct or indirect detection of the dark matter particle
is indispensable. The HESS source J1745-290 at the centre of our
galaxy (for the most recent anaysis see ref. \cite{Cembranos}) and 
the hints from the anomalous positron (or antiproton) abundance from the PAMELA mission
\cite{PAMELA} (see, for example, the analysis in ref. \cite{Cirelli})
suggest on the existence of the very heavy dark matter particle
when searching for the dark matter annihilation products.

MOND (with its few relativistic generalizations)
represents the alternative to the dark matter paradigm
(see a discussion in ref. \cite{MOND}).

Besides the supersymmetric, grand unified or extra dimensional
extensions of the SM, a very conservative alternative to the SM
was proposed in \cite{Palle1}, called the BY theory,
resolving the ultraviolet
singularity and the $SU(2)$ global anomaly problems.
Light and heavy Majorana neutrinos with flavour mixing and 
lepton CP violation could play a crucial role as hot and cold
dark matter particles in the evolution of the expanding and
rotating Universe \cite{Palle1,Palle2}.

The noncontractible space of the BY theory, as an alternative 
symmetry-breaking
mechanism to the Higgs one, introduces into the physical 
realm a new universal Lorentz and gauge invariant constant
(UV cutoff in the spacelike domain
of the Minkowski spacetime, see ref. \cite{Palle1}) 
$\Lambda=\frac{\hbar}{c d}=\frac{2}{g}\frac{\pi}{\sqrt{6}}
M_{W} \simeq 326 GeV$.

The enhanced strong coupling at small distances and the absence of
the asymptotic freedom in QCD are the immediate consequences of  
noncontractible space \cite{Palle3}.
We show that electroweak quantum loops with heavy
t-quark contributing to the CP violating processes of K and B mesons
are affected by the UV cutoff \cite{Palle4}.
The branching fraction for 
a rare decay $B_{s}\rightarrow \mu\mu$ is lower by more
than $30\%$ in the BY theory compared with the SM owing to the modified
short distance part of the amplitude \cite{Palle5}.

ATLAS and CMS experiments at the LHC reported recently \cite{CMS} a discovery
of the 125 GeV resonance.
It could be the Higgs boson of some theory beyond the SM, but it could be
also some pseudoscalar or scalar meson with a substantial component
of the pseudoscalar or scalar toponium \cite{Cea}.
Even the spin 1 boson cannot be excluded as an interpretation of the new 
125 GeV resonance \cite{Ralston}.

Anyhow, the Higgs mechanism
does not solve the mass problem of particles. Eventually, the solutions of
the coupled system of nonlinear integral Dyson-Schwinger equations
of the UV nonsingular BY theory could resolve the mass problem
of elementary particles \cite{Palle6}. 

In this paper, we study the implication of the UV cutoff 
to the leading QCD contribution for the forward-backward asymmetry
in the top-antitop production. The large discrepancy between the theory
and the experiment for this asymmetry observable is reported
by the Tevatron collaborations CDF and D0 \cite{Tevatron}.
Let us quote the most recent results of the CDF collaboration \cite{new}:
parton level asymmetry $A_{FB}(M_{t\bar{t}} < 450 GeV):\ Data (\pm stat 
\pm syst)= 0.084 \pm 0.046 \pm 0.026\ vs.\ SM\ expectation = 0.047 \pm 0.014$;      
$A_{FB}(M_{t\bar{t}} \geq 450 GeV):\ Data = 0.295 \pm 0.058 \pm 0.031\ vs.\
SM\ $ $expectation = 0.100 \pm 0.030$.
                                       
In the next chapter, we present the main ingredients of
the calculations while providing more details in the Appendix.
Results and Conclusions are given in the last chapter.

\section{Charge asymmetry at the parton level}

Almost invariably, various asymmetry observables of
the electroweak or strong interactions are very sensitive
to the details of the underlying processes.
It appears that the t-quark pair charge asymmetry can
test QCD loop corrections \cite{Kuehn}.
We shall study the dominant quark-antiquark annihilation
channel whose structure equals the electron-positron 
annihilation amplitude modulo coupling and gauge group constant
factors \cite{Berends1,Berends2,Berends3}.

Let us define the asymmetric part of the differential
cross sections \cite{Kuehn}

\begin{eqnarray*}
\frac{d \sigma^{q\bar{q}}_{A}}{d \cos \theta} =
 (\sigma^{q\bar{q}}_{A})'
\equiv \frac{1}{2}[\frac{d \sigma (q\bar{q} \rightarrow
 QX)}{d \cos \theta}-\frac{d \sigma (q\bar{q} \rightarrow
 \bar{Q}X)}{d \cos \theta}].
\end{eqnarray*}

Born cross section (symmetric part of the quark-antiquark
annihilation to leading order $\alpha^{2}_{s}$) is
given by \cite{Berends1,Kuehn}:

\begin{eqnarray*}
\frac{d \sigma(q\bar{q}\rightarrow Q\bar{Q};Born)}{d \cos \theta} =
\alpha^{2}_{s}\frac{T_{F}C_{F}}{N_{c}}
\frac{\pi \beta}{2 s} (1 + c^{2} +4 m^{2}), \\
T_{F}=\frac{1}{2}, C_{F}=\frac{4}{3}, N_{c}=3,
\beta=\sqrt{1-4 m^{2}}, m^{2}=\frac{m^{2}_{Q}}{s}, \\
s=E_{cm}^{2}, c=\beta \cos \theta,\ \angle(\vec{p}(q),\vec{p}(Q))=
\theta.
\end{eqnarray*}

The asymmetric part to the leading $\alpha^{3}_{s}$ order
consists of the virtual, soft and hard gluon emmission
differential cross sections
\cite{Kuehn,Berends2,Berends3}:

\begin{eqnarray}
(\sigma^{q\bar{q}}_{A})' &=& (\sigma^{q\bar{q}}_{A})'(virtual)
+ (\sigma^{q\bar{q}}_{A})'(soft) +
\int_{(I)} \frac{\partial^{4} (\sigma^{q\bar{q}}_{A}(hard)
-\sigma^{q\bar{q}}_{A}(soft))}{\partial \cos \theta
\partial \Omega_{\gamma} \partial k} 
d \Omega_{\gamma}d k   \nonumber \\
&+&\int_{(II)} \frac{\partial^{4} \sigma^{q\bar{q}}_{A}(hard)}
{\partial \cos \theta \partial \Omega_{\gamma} \partial k} 
d \Omega_{\gamma}d k,\nonumber \\
(I)&\ & 0 \leq k \leq k_{1},\ -1 \leq \cos \theta_{\gamma} \leq 1, 
\nonumber \\
(II)&\ & k_{1} \leq k \leq k_{2},\ 
g_{1}(k,E_{th}) \leq \cos \theta_{\gamma} \leq g_{2}(k,E_{th}).
\end{eqnarray}

The equations for the virtual, hard and soft gluon radiation
in the appendix of ref. \cite{Kuehn} are
obtained from the equations in \cite{Berends2,Berends3} in the limit
of the vanishing mass of incoming fermions.

The QCD in noncontractible space differs from the 
standard QCD when quantum loops are evaluated with
the cutoff in the spacelike domain.
Thus, one can find two possible sources of deviation from
the standard QCD calculation for the asymmetry function
$A^{\infty}(\cos\theta) = \sigma'_{A}/\sigma'_{Born}$:
(1) calculation of the running coupling $\alpha^{\Lambda}_{s}$
(see ref. \cite{Palle3}), (2) box diagram contribution
to the virtual correction \cite{Kuehn,Berends2,Berends3}:

\begin{eqnarray*}
(\sigma^{\Lambda}_{A})'&=&(\sigma^{\Lambda}_{A})'(virtual,\alpha^{\Lambda}_{s})
+(\sigma^{\Lambda}_{A})'(soft,\alpha^{\Lambda}_{s})
+(\sigma^{\Lambda}_{A})'(difference,\alpha^{\Lambda}_{s}) 
\end{eqnarray*}
\begin{eqnarray*}
&+&(\sigma^{\Lambda}_{A})'(hard,\alpha^{\Lambda}_{s})
=(\frac{\alpha^{\Lambda}_{s}}{\alpha^{\infty}_{s}})^{3}
(\sigma^{\infty}_{A})'(\alpha^{\infty}_{s})
+(\sigma^{\Lambda}_{A})'(virtual,\alpha^{\Lambda}_{s})
-(\sigma^{\infty}_{A})'(virtual,\alpha^{\Lambda}_{s}), 
\end{eqnarray*}
\begin{eqnarray*}
(\sigma^{\Lambda}_{Born})' = (\frac{\alpha^{\Lambda}_{s}}{\alpha^{\infty}_{s}})^{2}
(\sigma^{\infty}_{Born})',
\end{eqnarray*}
\begin{eqnarray*}
A^{\Lambda} = \frac{(\sigma^{\Lambda}_{A})'}{(\sigma^{\Lambda}_{Born})'},
\end{eqnarray*}
\begin{eqnarray}
A^{\Lambda}(\cos \theta)&=&A^{\infty}
+\delta A^{\Lambda}_{\alpha}+\delta A^{\Lambda}_{box}, 
\hspace{50 mm} \\
\delta A^{\Lambda}_{\alpha}&\equiv & \frac{\alpha_{s}^{\Lambda}-\alpha_{s}^{\infty}}
{\alpha_{s}^{\infty}}A^{\infty},\ 
\delta A^{\Lambda}_{box} \equiv
\frac{(\sigma^{\Lambda}_{A})'(virtual,\alpha_{s}^{\Lambda})-
(\sigma^{\infty}_{A})'(virtual,\alpha_{s}^{\Lambda})}
{(\sigma_{Born}^{\Lambda})'}, \nonumber
\end{eqnarray}
\begin{eqnarray*}
\Lambda\ denotes\ quantity\ in\ the\ BY\ theory,\ 
\infty\ denotes\ quantity\ in\ the\ SM. 
\end{eqnarray*}

We mean that $(\sigma^{\Lambda}_{A})'(virtual,\alpha_{s}^{\Lambda})$ 
is evaluated with $\alpha_{s}^{\Lambda}$ coupling, etc.
The calculation of the strong interaction running coupling
in noncontractible space was performed in the momentum
subtraction renormalization scheme to one loop order in ref.
\cite{Palle3}. Hard and soft gluon radiations do not contain
loop diagrams to leading $\alpha^{3}_{s}$ order.

Our main task should be a reevaluation of the interference
term in the cross section containing the box diagram in
the virtual correction term.
To accomplish this in noncontractible space, we have
to reduce the amplitude into pieces that are manifestly
translationally and Lorentz invariant.

We render light quark masses nonvanishing as a regulator 
of the collinear singularity that is canceled away in
the asymmetric cross sections. Infrared singularity is
controlled by the regulator gluon mass and is canceled
away in both $A^{\infty}$ and $\delta A_{box}^{\Lambda}$ asymmetry
parameters.
The virtual corrections can be represented with the
following expression \cite{Berends3}

\begin{eqnarray}
\frac{d \sigma_{A} (virtual)}{d \cos \theta}
=\alpha^{3}_{s}\frac{d^{2}_{abc}\beta_{t}}{32 N_{c}^{2} s}
[\sum_{j=1}^{7}w_{j}I_{j} - (\theta \rightarrow \pi-\theta)], \\
d^{2}_{abc}=\frac{40}{3},\ \beta_{t}=\sqrt{1-4 m_{t}^{2}/s}.
\nonumber
\end{eqnarray}

Definitions are given in the Appendix, as well as the procedure 
how to evaluate the integrals in noncontractible space to
maintain translational and Lorentz invariance.

Now we can compare the t-quark charge asymmetries to
the leading one loop order in the standard QCD and the
QCD in noncontractible space. The numerics and discussion can
be found in the last chapter.

\section{Results and conclusions}

The difference between the t-quark charge asymmetries 
of the standard QCD and the QCD in noncontractible space
lies in the additional two terms of Eq.(2) $\delta A_{\alpha}
^{\Lambda}$
and $\delta A_{box}^{\Lambda}$.
The first additional term $\delta A_{\alpha}^{\Lambda}$
can be evaluated using Table 1 derived from the formulae for $\alpha_{s}^{\Lambda}$
in ref. \cite{Palle3}. This correction can enhance the SM
asymmetry by up to $47\%$ for the largest parton $E_{cm}=14 TeV$.
The strong coupling $\alpha_{s}^{\Lambda}(\mu)$ is frozen at
$\mu \simeq 0.5 TeV$.

This is not enough to explain the asymmetry observed at the Tevatron
\cite{Tevatron}. Fortunately, the second additional term 
$\delta A_{box}^{\Lambda}$ provides the necessary enhancement
(see Figure 1 and Table 2).

We define the integrated charge asymmetry parameter as \cite{Kuehn}

\begin{eqnarray}
A_{int} \equiv \frac{\int_{0}^{1} \sigma_{A}'d \cos\theta}
{\int_{0}^{1} \sigma_{Born}'d \cos\theta}.
\end{eqnarray}

One can conclude that the charge asymmetries at the parton level
are enhanced in the BY theory by more than a factor of 3 (5) 
for Tevatron (LHC) center of mass energies. It is evident from 
Tables 1 and 2 that
the deviation from the SM is larger for higher $E_{cm}$ and
the virtual correction (box diagram) $\delta A_{box}^{\Lambda}$
dominates over the strong coupling correction $\delta A_{\alpha}
^{\Lambda}$. It means that the box diagram explores the deep
spacelike domain of the Minkowski spacetime to which the
asymmetry observable is very sensitive and, in addition,
there is no new negative compensation of the real hard and
soft contributions (no quantum loops to this order of perturbation) except
the new ${\alpha}^{\Lambda}_{s}$ factor.

\begin{table}
\caption{Running strong couplings at the scale $\mu = E_{cm}/2$
assuming $m_{u}=2.5 MeV$, $m_{d}=5.0 MeV$, $m_{s}=100 MeV$,
$m_{c}=1.6 GeV$, $m_{b}=4.8 GeV$, $m_{t}=172 GeV$ and
$\alpha_{s}(\mu=M_{Z})=0.12$.}
\hspace{20 mm}
\begin{tabular}{| c || c | c | c | c |} \hline
$E_{cm}(TeV)$ & 0.4 & 1.96 &  8 & 14  \\  \hline \hline
$\alpha_{s}^{\infty}(\mu)$ & 0.1077 & 0.08985 & 0.07886 & 0.0749 \\  \hline 
$\alpha_{s}^{\Lambda}(\mu)$& 0.1104  & 0.110  & 0.110  & 0.110\\  \hline
$\frac{\alpha_{s}^{\Lambda}-\alpha_{s}^{\infty}}{\alpha_{s}^{\infty}}
(\mu)$ &  0.0248  &  0.225 & 0.397 & 0.471 \\  \hline
\end{tabular}
\end{table}

\begin{table}
\caption{Integrated t-quark charge asymmetries for parton $E_{cm}$
evaluated with $E_{th}=0.9\times E_{cm}/2$ and $m_{t}=172 GeV$.}
\hspace{20 mm}
\begin{tabular}{| c || c | c | c | c |} \hline
$E_{cm}(TeV)$ & 0.4 & 1.96 &  8 & 14  \\  \hline \hline
$A^{\infty}_{int}$ & 0.0740 & 0.1774 & 0.1519 & 0.1449 \\  \hline 
$A^{\Lambda}_{int}$& 0.0939  & 0.661  & 0.805  & 0.874 \\  \hline
$A^{\Lambda}_{int}/A^{\infty}_{int}$
 &  1.27  &  3.73 & 5.30 & 6.03 \\  \hline
\end{tabular}
\end{table}

\epsfig{figure=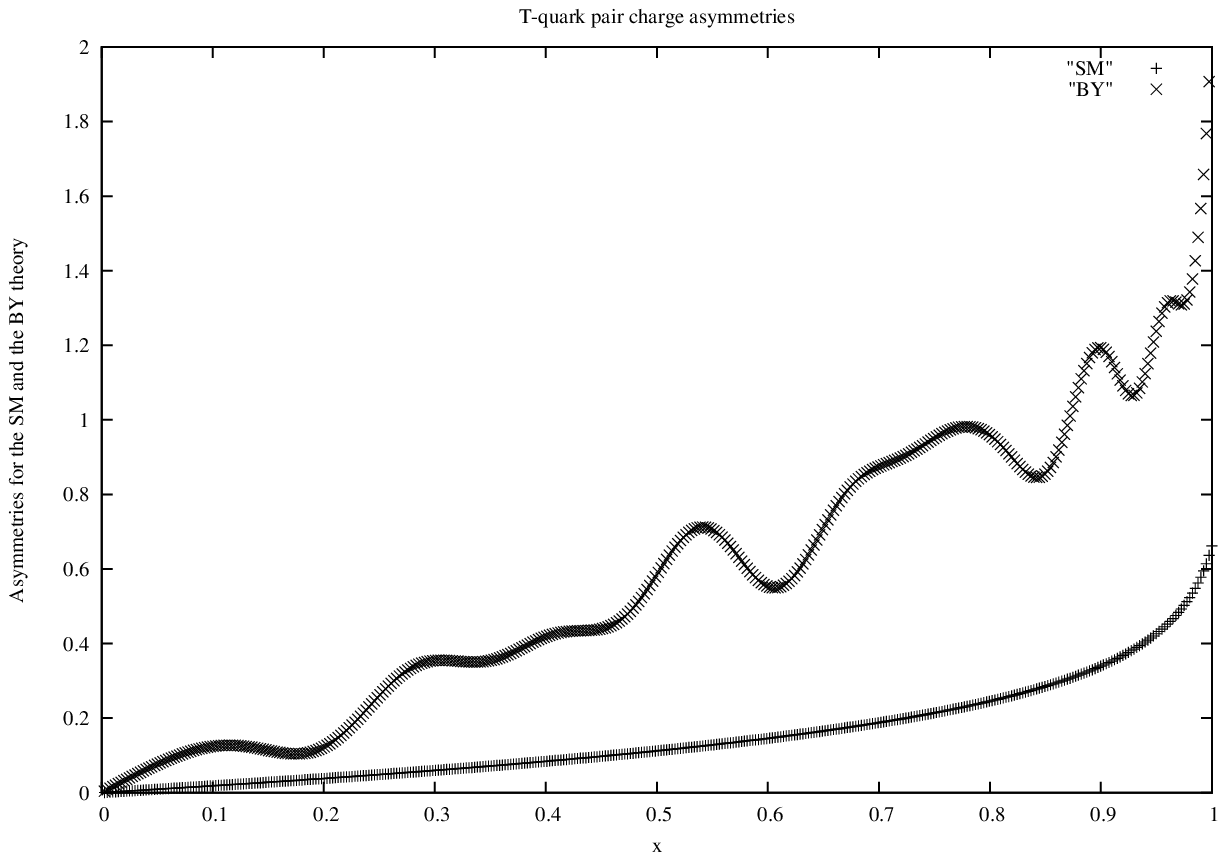, height=100 mm, width=130 mm}


{\bf Fig. 1: Asymmetry parameters $A^{\infty}$ and $A^{\Lambda}$ as 
a function of $x=\cos \theta$; \\ \hspace{40 mm}
$\ \ \ \ \ \ \ E_{cm}$=1.96 TeV, $m_{t}$=172 GeV, $E_{th}$=0.9 TeV}
\newline

To find charge asymmetry for hadrons, one has to convolve
parton cross sections with parton distributions. It is necessary
to solve DGLAP and BFKL equations in noncontractible space.
This work remains for the future.
It is very unlikely that higher orders of perturbation in the
strong coupling or new parton distributions can remove
large deviation of the asymmetry from the standard QCD found
at the parton level. If the LHC confirms the Tevatron
results, it will be necessary to investigate the issue to
higher perturbative order to reach higher accuracy, because 
to date, it is the largest discrepancy observed between
the standard QCD and the experiment.
\newline
\newline
{\bf Acknowledgment}
\newline
It is my pleasure to thank the referee for suggestions
that improve my presentation in the manuscript.
\newline

{\bf Appendix}
\newline
\newline
Since the details for the QCD running coupling evaluations can
be found in ref. \cite{Palle3} and the equations for the
SM asymmetries in ref. \cite{Kuehn}, in the Appendix, we outline
the equations for the virtual corrections in the SM and the BY theory
using notations of refs. \cite{Berends2,Berends3}.

Let us define the energy unit $E=E_{cm}/2$ and the dimensionless
mass of the light quark by $\overline{m}_{u}=m_{u}/E$ and the
t-quark by $\overline{m}_{t}=m_{t}/E$ \cite{Berends3}.
With previously defined $c=\beta_{t}\cos\theta$
the coefficients $w_{j}$ in the sum $\sum_{j=1}^{7}{w_{j}}I_{j}$
of Eq.(3) are as follows \cite{Berends3}:

\begin{eqnarray*}
w_{1}&=&1+c^{2}-2c^{3}+(1-2c)(\overline{m}_{u}^{2}+\overline{m}_{t}^{2}),\ 
w_{2}=2c(1-c)-\overline{m}_{u}^{2}-c\overline{m}_{t}^{2}, \\
w_{3}&=&2c(1-c)-\overline{m}_{t}^{2}-c\overline{m}_{u}^{2},\ 
w_{4}=2-c+c^{2}+\overline{m}_{t}^{2}+\overline{m}_{u}^{2}, \\
w_{5}&=&-1-c,\ w_{6}=1,\ w_{7}=1-c .
\end{eqnarray*}

We need further definitions to describe
the process $q(p_{+})+\bar{q}(p_{-})\rightarrow
t(q_{+})+\bar{t}(q_{-})$ and its amplitude
  \cite{Berends2,Berends3}:

\begin{eqnarray*}
P=\frac{1}{2}(p_{+}+p_{-}),\ \Delta=\frac{1}{2}(p_{+}-p_{-}),
\ Q=\frac{1}{2}(q_{+}-q_{-}), \\
\overline{\int}(f(k))\equiv\frac{4}{\pi^{2}}\Im \int\frac{f(k)d^{4}k}
{(\Delta)(Q)(+)(-)}, \\
(\Delta)=k^{2}-2 k\cdot\Delta-P^{2}+\imath\varepsilon,\ 
(Q)=k^{2}-2 k\cdot Q-P^{2}+\imath\varepsilon, \\
(\pm)=k^{2}\pm 2 k\cdot P+P^{2}-m_{gluon}^{2}+\imath\varepsilon.
\end{eqnarray*}

Now we can define dimensionless integrals $I_{j}$ of eq.(3):

\begin{eqnarray*}
I_{1}&=&E^{4}\overline{\int}(1),\ I_{2}=E^{2}\overline{\int}(k\cdot \Delta),
\ I_{3}=E^{2}\overline{\int}(k\cdot Q),
\ I_{4}=E^{2}\overline{\int}(k^{2}), \\
\ I_{5}&=&\overline{\int}((k\cdot P)^{2}), 
\ I_{6}=\overline{\int}((k\cdot \Delta)^{2}+(k\cdot Q)^{2}),
\ I_{7}=\overline{\int}((k\cdot \Delta)(k\cdot Q)).
\end{eqnarray*}

These integrals can be evaluated by the integrals from ref. \cite{Berends2}

\begin{eqnarray*}
[J;J_{\mu};J_{\mu\nu}]=\int d^{4}k\frac{[1;k_{\mu};k_{\mu}k_{\nu}]}
{(\Delta)(Q)(+)(-)},
\end{eqnarray*}

that are expressed 
in terms of nine functions: $F,G,F_{\Delta},F_{Q},G_{\Delta},G_{Q},
H_{P},$ $H_{\Delta},H_{Q}$.

Let us represent seven integrals $I_{j}$ of ref. \cite{Berends3}
in terms of functions from \cite{Berends2}:

\begin{eqnarray}
I_{1}&=&\frac{4}{\pi^{2}}\frac{F+G}{2 P^{2}}E^{4},  \nonumber \\
I_{2}&=&\frac{4}{\pi^{2}}(\Delta^{2}J_{\Delta}+\Delta\cdot Q J_{Q}) 
E^{2},  \nonumber \\
I_{3}&=&\frac{4}{\pi^{2}}(Q^{2}J_{Q}+\Delta\cdot Q J_{\Delta})E^{2},  \nonumber \\
I_{4}&=&\frac{4}{\pi^{2}}(4 K_{O}+P^{2}K_{P}+\Delta^{2}K_{\Delta}
+Q^{2}K_{Q}+2 \Delta\cdot Q K_{X})E^{2},  \\
I_{5}&=&\frac{4}{\pi^{2}}(K_{O}P^{2}+K_{P}(P^{2})^{2}), \nonumber \\
I_{6}&=&\frac{4}{\pi^{2}}(K_{O}(\Delta^{2}+Q^{2})
+K_{\Delta}((\Delta^{2})^{2}+(\Delta\cdot Q)^{2})
+K_{Q}((\Delta\cdot Q)^{2}+(Q^{2})^{2}) \nonumber \\
&+& 2 K_{X}\Delta\cdot Q (\Delta^{2}+Q^{2})), \nonumber \\
I_{7}&=&\frac{4}{\pi^{2}}(\Delta\cdot Q K_{O}
+\Delta\cdot Q \Delta^{2} K_{\Delta}
+\Delta\cdot Q Q^{2} K_{Q}
+ K_{X}(Q^{2}\Delta^{2}+(\Delta\cdot Q)^{2})), \nonumber
\end{eqnarray}

where $J_{\Delta},\ J_{Q},\ K_{O},\ K_{P},\ K_{\Delta},
\ K_{Q}\ and\ K_{X}$ functions are defined in terms of
nine functions $F,...,H_{Q}$ \cite{Berends2}.

We use standard definitions for the scalar two, three and
 four point functions \cite{vanOld}:

\begin{eqnarray*}
B_{0}(p;m_{1},m_{2})=(\imath \pi^{2})^{-1}
\int d^{4} k [k^{2}-m_{1}^{2}+\imath\epsilon]^{-1}
[(k+p)^{2}-m_{1}^{2}+\imath\epsilon]^{-1},\\
C_{0}(p_{1},p_{2};m_{0},m_{1},m_{2})=
(\imath \pi^{2})^{-1}\int d^{4} k[k^{2}-m_{0}^{2}+\imath\epsilon]^{-1}
[(k+p_{1})^{2}-m_{1}^{2}+\imath\epsilon]^{-1} \\
\times[(k+p_{2})^{2}-m_{2}^{2}+\imath\epsilon]^{-1},
\end{eqnarray*}
\begin{eqnarray*}
D_{0}(p_{1},p_{2},p_{3};m_{0},m_{1},m_{2},m_{3})=
(\imath \pi^{2})^{-1}\int d^{4} k[k^{2}-m_{0}^{2}+\imath\epsilon]^{-1} \\
\times[(k+p_{1})^{2}-m_{1}^{2}+\imath\epsilon]^{-1} 
[(k+p_{2})^{2}-m_{2}^{2}+\imath\epsilon]^{-1}
[(k+p_{3})^{2}-m_{3}^{2}+\imath\epsilon]^{-1}.
\end{eqnarray*}

In ref. \cite{Berends2} expressions for all nine 
functions $F,...,H_{Q}$ in the standard QCD can be found.
The same functions
have to be expressed by the previous scalar two, three and four point
Green functions in noncontractible space in order to properly 
restore translational invariance \cite{Palle3,Palle4,Palle5}.

Functions $G,F_{\Delta},F_{Q}$ have already a suitable form
of the three point functions \cite{Berends2}:

\begin{eqnarray*}
G &=& \int d^{4} k (\Delta)^{-1}(Q)^{-1}(+)^{-1},\ 
F_{\Delta} = \int d^{4} k(\Delta)^{-1}(+)^{-1}(-)^{-1},\ \\
F_{Q} &=& \int d^{4} k(Q)^{-1}(+)^{-1}(-)^{-1}.
\end{eqnarray*}

Note that
all expressions in ref. \cite{Berends2} are derived under the
assumption of \newline $m_{gluon}\equiv\lambda \ll m_{u},m_{t},E_{cm}$.
From their definitions, $G_{\Delta}$ and 
$G_{Q}$ can be expressed as:

\begin{eqnarray}
\Im G_{Q} = \frac{1}{\beta^{2}_{t}}\Im F_{Q}
+\frac{2 \pi^{2}}{s \beta^{2}_{t}}[\Re B_{0}(-2P;\lambda,\lambda)
-\Re B_{0}(-Q-P;\lambda,m_{t})], \nonumber \\
\Im G_{\Delta} = \frac{1}{\beta^{2}_{u}}\Im F_{\Delta}
+\frac{2 \pi^{2}}{s \beta^{2}_{u}}[\Re B_{0}(-2P;\lambda,\lambda)
-\Re B_{0}(-\Delta-P;\lambda,m_{u})].
\end{eqnarray}

For functions $F,H_{P},H_{\Delta},H_{Q}$, we derive the equations
that allow to put these functions in the alternative form
expressed only through scalar n-point integrals.

The linear system for the $F$ function looks as

\begin{eqnarray}
p_{1}^{2}\eta_{1}+p_{1}\cdot p_{2}\eta_{2}+p_{1}\cdot p_{3}\eta_{3}
=R_{1}, \nonumber \\
p_{1}\cdot p_{2}\eta_{1}+p_{2}^{2}\eta_{2}+p_{2}\cdot p_{3}\eta_{3}
=R_{2}, \nonumber \\
p_{1}\cdot p_{3}\eta_{1}+p_{2}\cdot p_{3}\eta_{2}+p_{3}^{2}\eta_{3}
=R_{3}, 
\end{eqnarray}
\begin{eqnarray*}
R_{1}=\frac{1}{2}[\Re C_{0}(p_{2},p_{3};m_{0},m_{2},m_{3})
-\Re C_{0}(p_{2}-p_{1},p_{3}-p_{1};m_{1},m_{2},m_{3}) \\
-(p_{1}^{2}-m_{1}^{2}+m_{0}^{2})
\Re D_{0}(p_{1},p_{2},p_{3};m_{0},m_{1},m_{2},m_{3})], \\
R_{2}=\frac{1}{2}[\Re C_{0}(p_{1},p_{3};m_{0},m_{1},m_{3})
-\Re C_{0}(p_{2}-p_{1},p_{3}-p_{1};m_{1},m_{2},m_{3}) \\
-(p_{2}^{2}-m_{2}^{2}+m_{0}^{2})
\Re D_{0}(p_{1},p_{2},p_{3};m_{0},m_{1},m_{2},m_{3})], \\
R_{3}=\frac{1}{2}[\Re C_{0}(p_{1},p_{2};m_{0},m_{1},m_{2})
-\Re C_{0}(p_{2}-p_{1},p_{3}-p_{1};m_{1},m_{2},m_{3}) \\
-(p_{3}^{2}-m_{3}^{2}+m_{0}^{2})
\Re D_{0}(p_{1},p_{2},p_{3};m_{0},m_{1},m_{2},m_{3})], \\
p_{1}=2 P,\ p_{2}=P-\Delta,\ p_{3}=P-Q,\ 
m_{0}=m_{1}=\lambda,\ m_{2}=m_{u},\ m_{3}=m_{t} \\
\Rightarrow \Im F = -\Im F_{Q}+2\pi^{2}(\Delta^{2}\eta_{2}
+\Delta\cdot Q \eta_{3}).
\end{eqnarray*}

Similarly, we derive the linear system for $H$ functions

\begin{eqnarray}
p_{1}^{2}\rho_{1}+p_{1}\cdot p_{2}\rho_{2}= M_{1}, \nonumber \\
p_{1}\cdot p_{2}\rho_{1}+p_{2}^{2}\rho_{2}= M_{2},
\end{eqnarray}
\begin{eqnarray*}
M_{1}&=&\frac{1}{2}[\Re B_{0}(p_{2};\lambda,m_{2})-\Re B_{0}(p_{2}-p_{1};m_{1},m_{2})
+(-\lambda^{2}+m_{1}^{2}-p_{1}^{2}) \\
 &\times&  \Re C_{0}(p_{1},p_{2};\lambda,m_{1},m_{2})], \\
M_{2}&=&\frac{1}{2}[\Re B_{0}(p_{1};\lambda,m_{1})-\Re B_{0}(p_{2}-p_{1};m_{1},m_{2})
+(-\lambda^{2}+m_{2}^{2}-p_{2}^{2}) \\
 &\times& \Re C_{0}(p_{1},p_{2};\lambda,m_{1},m_{2})], \\
p_{1}&=&P-\Delta,\ p_{2}=P-Q,\ m_{1}=m_{u},\ m_{2}=m_{t}\\
& & \Rightarrow \Im H_{P}=\Im G+\pi^{2}(\rho_{1}+\rho_{2}),\ 
\Im H_{\Delta}=-\pi^{2}\rho_{1},\ \Im H_{Q}=-\pi^{2}\rho_{2}.
\end{eqnarray*}

The validity of new forms for $F,H_{P},H_{\Delta}\ and\ H_{Q}$ is
also checked numerically.

The virtual corrections can be evaluated by eq. (A.1) of
ref. \cite{Berends3} or by eq. (12) of ref.\cite{Berends2}.

We are now prepared for the crucial step to calculate virtual
corrections in noncontractible space defining scalar
n-point integrals in noncontractible space. $B_{0}^{\Lambda}$
function is outlined in refs. \cite{Palle3,Palle4,Palle5}.
The similar procedure should be applied to the three point function:  

\begin{eqnarray}
\Re C_{0}^{\infty} &=& \Re C_{0}^{\Lambda} + \delta C_{0}^{\Lambda}(symm), \\
\delta C_{0}^{\Lambda}(symm)&=&\frac{1}{3}
[\delta C_{0}^{\Lambda}(p_{1},p_{2};m_{0},m_{1},m_{2})
+\delta C_{0}^{\Lambda}(-p_{1},p_{2}-p_{1};m_{1},m_{0},m_{2})\nonumber  \\
& & +\delta C_{0}^{\Lambda}(-p_{2},p_{1}-p_{2};m_{2},m_{0},m_{1})], \nonumber
\end{eqnarray}
\begin{eqnarray*}
\delta C_{0}^{\Lambda}(p_{1},p_{2};m_{0},m_{1},m_{2})
=\pi^{-2}\int^{1/\Lambda}_{0} dw w^{-5} \int^{+1}_{-1}dx
\sqrt{1-x^{2}}\int^{+1}_{-1}dy  \\ \times \int^{2\pi}_{0}d\phi 
[-k^{2}-m_{0}^{2}]^{-1} 
[-k^{2}+2 (k \cdot p_{1})+p_{1}^{2}-m_{1}^{2}]^{-1} \\ \times
[-k^{2}+2 (k \cdot p_{2})+p_{2}^{2}-m_{2}^{2}]^{-1}(k=w^{-1}),\\
where\ (k \cdot p_{1})=\imath kx(p_{1})^{0}-\vec{k}\cdot\vec{p}_{1}, \\
\vec{k}=k\sqrt{1-x^{2}}(\sqrt{1-y^{2}}\cos \phi,
\ \sqrt{1-y^{2}}\sin \phi,\ y).
\end{eqnarray*}

All the imaginary parts of the subintegral function in
$\delta C_{0}^{\Lambda}$ are erased by integration
as odd functions in variable x.

The same decomposition is possible for the four point function,
although with four terms necessary for symmetrization in
$\delta D_{0}^{\Lambda}$.

Multidimensional numerical integrations in virtual and real
gluon radiations are performed
by Suave routine from CUBA library \cite{Hahn} to the
relative accuracy of ${\cal O}(10^{-4})$ with up to
50 million of sampling points per integral.

\end{document}